\documentstyle[12pt]{article}
\renewcommand{\theequation}{\arabic {section}.\arabic{equation}}
\renewcommand{\thefootnote}{\fnsymbol{footnote}}

\newcommand{\EQ}{\begin{equation}}
\newcommand{\EN}{\end{equation}}
\newcommand{\bea}{\begin{eqnarray}}
\newcommand{\ena}{\end{eqnarray}}
\newcommand{\vs}[1]{\vspace{#1 mm}}

\newcommand{\uda}{\nearrow \kern-1em \searrow}

\newcommand{\PL}[1]{Phys.\ Lett.\ {\bf #1}}

\newcommand{\PR}[1]{Phys.\ Rev.\ {\bf #1}}

\newcommand{\PTP}[1]{Prog.\ Theor.\ Phys.\ {\bf #1}}

\newcommand{\AJ}[1]{Astorophys. \ J.\ {\bf #1}}

\begin{document}
\setlength{\baselineskip}{7mm}

\begin{titlepage}
\setcounter{page}{0}
\begin{flushright}
EPHOU 99-007\\
OU-HET-319   \\
May 1999
\end{flushright}
\vs{6}
\begin{center}
{\Large Analytic Solutions of Teukolsky Equation in Kerr-de Sitter and 
Kerr-Newman-de Sitter Geometries}

\vs{5}
{\bf {\sc
Hisao Suzuki\footnote{e-mail address:
hsuzuki@particle.sci.hokudai.ac.jp}}
\\
{\em Department of Physics, \\
Hokkaido
University \\  Sapporo, Hokkaido 060 Japan} \\
{\sc Eiichi Takasugi\footnote{
e-mail address: takasugi@het.phys.sci.osaka-u.ac.jp}
 and  Hiroshi Umetsu\footnote{
e-mail address: umetsu@het.phys.sci.osaka-u.ac.jp}}\\
{\em Department of Physics,\\
Osaka University \\
Toyonaka, Osaka 560 Japan}}\\
\end{center}
\vs{6}

\begin{abstract}
The analytic solution of Teukolsky equation in Kerr-de Sitter 
and Kerr-Newman-de Sitter geometries is presented  and the 
properties of the solution are examined. In particular, 
we show that our solution satisfies the Teukolsky-Starobinsky 
identities explicitly and fix the relative normalization 
between solutions with the spin weight $s$ and $-s$.   
\end{abstract}
\end{titlepage}

\renewcommand{\thefootnote}{\arabic{footnote}}
\setcounter{footnote}{0}


\section{Introduction}
 
In our series of papers, Mano, Suzuki and Takasugi~[1, 2] 
have constructed the analytic solution of the perturbation 
equation of massless fields in Kerr geometries 
which is commonly called Teukolsky equation~[3]. 
The solution enabled us to investigate various 
properties of black hole analytically~[4],  
the scattering problem of particle emitted in black hole,  
the analytical expression for the absorption rate 
by Kerr black hole which 
was one of the main concern by Chandrasekhar~[5], and 
the Teukolsky-Starobinsky identities. 

Our solution is expressed in the form of series of hypergeometric 
functions and Coulomb wave functions. The coefficients of 
these series are determined by solving three term recurrence 
relations. Two series have different convergence regions 
and they are matched in the region where both series converge. 
This method turns out to be quite powerful 
in a practical calculation
and is sucessfully applied to examine  gravitational waves 
from a binary in a post-Newtonian expansion; to construct 
the template of the gravitational wave emitted from a particle 
moving around a Kerr black hole~[6] and 
the rate of the gravitational wave absorbed by the black hole~[7]. 

In our previous paper~[8], Suzuki, Takasugi and Umetsu have 
extended this method to solve the perturbation equations of 
massless fields (the Teukolsky equation) 
in Kerr-de Sitter and Kerr-Newman-de Sitter geometries. We 
found the transformations such that both the angular 
and the radial equations 
are reduced to the Heun's equation~[9]. The solution of 
Heun's equation is expressed in the form of series 
of hypergeometric functions and its coefficients are 
determined by three term recurrence relations, similarly to 
the Kerr geometry case.  It should be noted that electromagnetic 
field and gravitational fields couple each other 
in Kerr-Newman-de Sitter geometries and 
do not follow the equation we considered there, although the fields follow 
the equation in Kerr-de Sitter geometries.  

The solution of the radial Teukolsky equation given in Ref.[8] 
is convergent  
for $r<r_+'$ with $r_+'$ being the de Sitter horizon. 
In this paper, we give  the solution  valid 
in the entire physical region. This is achieved 
by constructing another solution which is convergent 
for $r_+<r$ and then by matching this solution and 
the previous soluiton in the reigion where both 
solutions are convergent.  
Then, we examine porperties of the solution in detail. In 
particular, we show analytically that our solution satisfies 
the Teukolsky-Statrobinsky (T-S) identities~[10], if we properly take 
the relative normalizaition between the solution with the spin 
weight $s$ and the one with $-s$. 

It is quite surprising that the analytic solution is 
obtained for Kerr-de Sitter and Kerr-Newman-de Sitter 
geometies, similarly to the Kerr geometry case. 
This might be the reflection of an underlying symmetry 
which the gravitational theory contains. 
Although the use of this solution is  
less clear in comparison with that in the Kerr geometries, 
we believe that our finding is important not only because 
the technique we used can be exteded to other phenomena, 
but also because the solution may become relevant to 
the early stage of universe where the cosmological constant
play important role.

In Sec.2, we review the transformation of the radial equation 
to Heun's equation, because we change some definitions of 
parameters from the previous paper[8]. In Sec.3, the solution 
is given in the form of series of hypergeometric functions. 
In particular, we present the solution which is convergent 
around the de Sitter horizon. We discuss the convergence 
region of the solution. Then, we match two solutions in the 
region where both solutions converge. In Sec.4, we explicitly 
show that our solution satisfies the Teukolsky-Starobinsky 
identities and the relative normalization between the solutions 
with the spin weight $s$ and $-s$. We found some identities 
involving coefficients of series. Summary and discussions are 
given in Sec.5.


\section{Teukolsky equation for the Kerr(-Newman)-de Sitter geometry 
and Heun's equation}
\setcounter{equation}{0}

We consider the Teukolsky equations for the Kerr-Newman-de Sitter 
geometries.  
In the Boyer-Lindquist coordinates the Kerr-Newman-de Sitter metric
has 
the form,
\begin{eqnarray}
ds^2 &=& -\rho^2
\left(\frac{dr^2}{\Delta_r}+\frac{d\theta^2}{\Delta_\theta}\right)
-\frac{\Delta_\theta \sin^2\theta}{(1+\alpha)^2 \rho^2}
[adt-(r^2+a^2)d\varphi]^2 \nonumber \\ 
&&  +\frac{\Delta_r}{(1+\alpha)^2 \rho^2}(dt-a\sin^2\theta
d\varphi)^2,
\end{eqnarray}
where 
\begin{equation}
\begin{array}{cc}
\multicolumn{2}{c}{{\displaystyle 
\Delta_r=(r^2+a^2)\left(1-\frac{\alpha}{a^2}r^2\right)-2Mr+Q^2
=-\frac{\alpha}{a^2}(r-r_+)(r-r_-)(r-r'_+)(r-r'_-),}} \\
\Delta_\theta=1+\alpha\cos^2\theta, &
{\displaystyle \alpha=\frac{\Lambda a^2}{3}}, \\
\bar{\rho}=r+ia\cos\theta, &
\rho^2=\bar{\rho}\bar{\rho}^*. \\
\end{array}
\end{equation}
Here $\Lambda$ is the cosmological constant, $M$ is the mass of the
black hole, $aM$ its angular momentum and $Q$ its charge.

We assume that the time and the azimuthal angle dependence of the 
field has the form $e^{-i(\omega t-m\phi)}$. Then, the equation for 
the radial part with spin $s$ and charge $e$ is  
given by 
\begin{eqnarray}
&& \left[
\Delta_r {\cal D}_1 {\cal D}_s^\dag +2(1+\alpha)(2s-1)i\omega r 
-\frac{2\alpha}{a^2}(s-1)(2s-1) \right.  \nonumber \\
&& \makebox[5mm]{}  \left.  
+\frac{-2(1+\alpha)eQKr+iseQr\partial_r \Delta_r+e^2Q^2r^2}
{\Delta_r}
- 2iseQ-s(1-\alpha)-\lambda_s \right]R_s = 0\;, 
\nonumber\\
\label{eqn:Rs}
\end{eqnarray}
where 
\begin{eqnarray}
{\cal D}_n&=&\partial_r-\frac{i(1+\alpha)K}{\Delta_r}
+n\frac{\partial_r \Delta_r}{\Delta_r}, \nonumber \\
{\cal D}_n^\dag&=&\partial_r+\frac{i(1+\alpha)K}{\Delta_r}
+n\frac{\partial_r \Delta_r}{\Delta_r},   
\end{eqnarray}
with $K=\omega(r^2+a^2)-am$. In our previous paper~[8], 
we showed that solutions are characterized by the 
characteristic exponent (shifted angular momentum) $\nu$. 

In this paper, we use slightly different definitions of the 
separation constant $\lambda_s$ and the characteristic exponents 
$\nu$ and $-\nu-\omega$, where $\omega$ is the parameter appearing in  
the Heun's equation, is given explicitly in Eq.(2.17) and 
is different from the angular frequency in Eq.(2.3). 
We redefined $\lambda_s$ and $\nu$ as 
\bea
\lambda_s &\to& \lambda_s +s(1-\alpha)\;,\nonumber\\
\nu&\to& \nu-\frac{\omega-1}{2} \;.
\ena
With these modifications, as we showed in Ref.[8], $\lambda_s$ becomes 
an even function of spin variable $s$, 
\bea
\lambda_s=\lambda_{-s}\;,
\ena
and if $\nu$ is a characteristic exponent, then 
$-\nu-1$  becomes the one. That is, both 
\bea
\nu\; \;\;{\rm and }\;\;-\nu-1\;.
\ena
become characteristic exponents. 

With a new $\lambda_s$, the radial Teukolsky equation is explicitly 
written by 
\begin{eqnarray}
&&\makebox[-10mm]{}\Bigg\{ \ 
\Delta_r^{-s}\frac{d}{dr}\Delta_r^{s+1}\frac{d}{dr}
+\frac{1}{\Delta_r}\left[ (1+\alpha)^2 \left(K-\frac{eQr}{1+\alpha}
\right)^2
- is(1+\alpha)\left(K-\frac{eQr}{1+\alpha}\right) \frac{d\Delta_r}
{dr} \right] \nonumber \\
&& \makebox[-5mm]{}
+4is(1+\alpha)\omega r -\frac{2\alpha}{a^2}(s+1)(2s+1) r^2
+s(1-\alpha)-2iseQ -\lambda_s \ \Bigg\} R_s = 0. \label{eqn:Rr}
\end{eqnarray}
This equation has five regular singularities at  $r_\pm, r'_\pm$ and 
$\infty$ which are assigned such that  
$r_\pm \rightarrow M\pm\sqrt{M^2-a^2-Q^2}$ and 
$\displaystyle{r'_\pm \rightarrow \pm\frac{a}{\sqrt{\alpha}}}$ 
in the limit $\alpha \rightarrow 0$ $(\Lambda \rightarrow 0)$. 
 
Next, we use a variable $x$ rather than $z$ used  
in our previous paper~[8], which are defined by
\begin{eqnarray}
x &=& 1-z=\frac{(r_- -r'_-)}{(r_- -r_+)}\frac{(r-r_+)}{(r-r'_-)}.
\end{eqnarray}
This transformation from $r$ to $x$ maps 
the inner horizon $r_-$, the outer horizon $r_+$, the de Sitter
horizon 
$r'_+$, $r'_-$ and $\infty$ to 0, 1, $x_r$, $\infty$ and $x_{\infty}$, 
respectively,
\bea
 x_r &=& 1-z_r=\frac{(r_- -r'_-)}{(r_- -r_+)}
 \frac{(r'_+-r_+)}{(r'_+-r'_-)},\nonumber \\
 x_{\infty} &=& 1-z_{\infty}
 =\frac{(r_- -r'_-)}{(r_- -r_+)}\;. 
\ena
In the followings, we omit the subscript $s$ for the radial wave 
function.

To proceed further, we define the following parameters,
\bea
A_{i\pm}&=&\frac{1}{2}\left\{-s \pm (2a_i+s  )\right\}\qquad 
(i=1,2,3,4), 
\ena
Here $a$'s are pure imaginary numbers defined by 
\bea
a_1&=&i\frac{a^2(1+\alpha)}{\alpha} \frac{ 
\left(\omega(r_+^2 + a^2)-am-\frac{eQr_+}{1+\alpha}\right)}
{ (r'_+ - r_+)(r'_- - r_+)(r_- - r_+)} \;,\nonumber\\
a_2&=&i\frac{a^2(1+\alpha)}{\alpha} \frac{ 
\left(\omega(r_-^2 + a^2)-am-\frac{eQr_-}{1+\alpha}\right)}
{ (r'_+ - r_-)(r'_- - r_-)(r_+ - r_-)}\;,
  \nonumber\\
a_3&=&i\frac{a^2(1+\alpha)}{\alpha} \frac{ 
\left(\omega({r'_+}^2 + a^2)-am-\frac{eQr'_+}
{1+\alpha}\right)}
{ (r_- - r'_+)(r'_- - r'_+)(r_+ - r'_+)} \;,\nonumber
\\
a_4&=&i\frac{a^2(1+\alpha)}{\alpha} \frac{
\left(\omega(r_-'^2 + a^2)-am-\frac{eQr_-'}{1+\alpha}\right)}
{(r_- - r'_-)(r'_+ - r'_-)(r_+ - r'_-)} 
\ena
and they satisfy the relation
\bea
a_1+a_2+a_3+a_4=0.
\ena

Now, following the transformation given in Ref.[8], 
we can factor out the singularity at $x_\infty$ by the 
transformation as 
\bea
R^{\nu}_{in;\{0,1\};s}(x) =
(-x)^{A_1}(1-x)^{A_2}\left(\frac{x-x_r}{1-x_r}\right)^{A_3}
\left(\frac{x-x_\infty}{1-x_\infty}\right)^{2s+1}
f ^{\nu}_{\{0,1\};s}(x)\;, 
\ena
where we choose $A_1=A_{1-}$ for the solution to satisfy the 
incoming boundary condition at the outer horizon, and 
the other $A_i (i=2,3)$ takes either $A_{i+}$  or $A_{i-}$. 
Then, we find that $f^{\nu}_{\{0,1\}}(x)$ 
satisfies the following Heun's equation
\bea
\left\{ \ \frac{d^2}{dx^2} + \left[\frac{\gamma}{x} +
\frac{\delta}{x-1}
+\frac{\epsilon}{x-x_r} \right] \frac{d}{dx} 
+ \frac{\sigma_+ \sigma_-x + v}{x (x-1)(x-x_r)} \ \right\}
f^{\nu}_{\{0,1\};s}(x)  =0\;,
\ena
where 
\bea
\gamma=2A_1+s+1, \quad && \delta=2A_2+s+1, \quad \epsilon=2A_3+s+1\;, 
\nonumber\\
\sigma_\pm \makebox[-3mm]{}&=&\makebox[-3mm]{} A_1 + A_2 + A_3 +
A_{4\mp}
+2s+1 \;.
\ena
Parameters $\gamma$, $\delta$, $\epsilon$ and $\sigma_{\pm}$ 
satisfy the following relation, 
\bea
\gamma+\delta-1=\sigma_++\sigma_--\epsilon \equiv \omega\;,
\ena
which is required for Eq.(2.15) to be a Heun's equation. 
It should be noted that the parameter $\omega$ defined 
above is a commonly used notation for Heun's equation and is different 
from the angular frequency. The remaining parameter $v$ is given 
by
\bea
v &=& -\frac{2a^4 (1+\alpha)^2}{\alpha^2 {\cal D}}
\frac{(r_- - r'_+)^2 (r_- - r'_-)^2 (r_+ - r'_-) (r'_+ - r'_-)}
{r_+ - r_-} \nonumber \\
&& \makebox[-3mm]{}
\Bigg\{-\omega^2 r_+^3(r_+ r_- - 2r_- r'_+ + r_+ r'_+)
+2a\omega(a\omega-m)r_+ (r_- r'_+ - r_+^2)  \nonumber\\
&& -a^2(a\omega-m)^2(2r_+ - r_- - r'_+)  
+\frac{eQ}{1+\alpha} 
\big[ \ \omega{r_+}^2(r_+ r_- + r_+^2 -3r_- r'_+ + r_+
r'_+)\nonumber\\
&&-a(a\omega-m)(r_+ r_- -3r_+^2 +r_- r'_+ + r_+ r'_+)
\  \big]  
+ \left(\frac{eQ}{1+\alpha}\right)^2 r_+(-r^2_+ +r_- r'_+) \
\Bigg\} \nonumber\\ 
&& \makebox[-3mm]{}
-\frac{2isa^2(1+\alpha)}{\alpha}
\frac{\left[ \omega (r_+ r'_- +a^2)-am 
-\frac{eQ}{1+\alpha}\frac{r_+ +r'_-}{2} \right]}
{(r_+ - r_-)(r'_+ - r'_-)(r_+ - r'_-)}  \nonumber \\
&& \makebox[-3mm]{} 
-(s+1)\left((2s+1)\left[\frac{2{r'_-}^2}
{(r_+ - r_-)(r'_+ -r'_-)}+x_\infty\right] 
+\left[ (1+x_r)A_1+x_rA_2+A_3 \right]\right)\nonumber\\
&& \makebox[-3mm]{}
-2A_1(x_r A_2+A_3)+\frac{a^2}{\alpha(r_+ - r_-)(r'_+ - r'_-)} 
\left[-\lambda_s+s(1-\alpha)\right]\;,
\ena
where ${\cal D}$ is the discriminant of $\Delta_r = 0$,
\bea
{\cal D}&=&(r_+ - r_-)^2 (r_+ - r'_+)^2 (r_+ - r'_-)^2
(r_- - r'_+)^2 (r_- - r'_-)^2 (r'_+ - r'_-)^2 \nonumber \\
&=& \frac{16a^{10}}{\alpha^5}\Bigg\{ \ 
(1-\alpha)^3\left[M^2-(1-\alpha)(a^2+Q^2)\right] 
-\frac{16\alpha^2}{a^4}(a^2+Q^2)^3 \nonumber \\ 
&&  +\frac{\alpha}{a^2}\left[-27M^4+36(1-\alpha)M^2(a^2+Q^2)
- 8(1-\alpha)^2(a^2+Q^2)^2\right]
\ \Bigg\}. 
\ena
A parameter $\omega$ in the definition of $v$ is an angular 
frequency. 
 
Modifications in Eq.(2.5) and the use of the variable 
$x$ rather than $z$ cause  changes of parameters in the 
Heun's equation from those in Ref.[8] . The changes of 
parameters are shown below:  
\bea
B_1 \;\to\; A_2,\;\; B_2 \;\to\; A_1,\;\; B_3 \;\to\; A_3,\;
\nonumber\\
(\alpha,\;\beta)\;\to\;(\sigma_+,\;\sigma_-),\;-q\;\to\; v, 
\;\;v\;\to -v-\sigma_+\sigma_-,\;\;a_H\;\to\;x_r\;. 
\ena
Parameters in the left-hand side are those used in Ref.[8] and 
those in the right-hand side are those used in this paper. 
The $v$ in Eq.(2.18) is derived from the corresponding parameter $v$ 
of Eq.(3.10) in Ref.[8] by exchanging $r_+$ and 
$r_-$ and changing  of $\lambda_s$  in Eq.(2.5).  
There was a miss-print in $v$ in our previous paper~[8] where 
$-\lambda_s-2iseQ$ should be read $-\lambda_s$. 

\section{Solutions of Teukolsky equation}
\setcounter{equation}{0}
In this section, we derive the solutions which are convergent 
in the region including the outer and the inner horizon and 
also the solutions which are convergent 
in the region including the de Sitter  horizon and $\infty$. 
Then, by matching these solutions in the region where both 
solutions are convergent, we obtain the solutions valid in the 
entire region. 
 
\vskip 3mm
\noindent
(A) Solutions convergent around outer horizon 

There are two independent solutions; one satisfies the 
incoming boundary condition at the outer horizon and 
the other does the outgoing boundary condition. 
They are expressed as series of hypergeometric functions 
and these series converges in the region, $r<r_+'$ with 
$r_+'$ being the de Sitter horizon. 

\vskip 3mm 
\noindent
{\it (a-1) The solution which satisfies the incoming boundary
condition 
at the outer horizon} 
 
The solution is given  by taking into account of 
changes in Eq.(2.5) and 
adopting the variable $x$ as
\bea
R^{\nu}_{in; \{0,1\};s}(x) =
(-x)^{A_1}(1-x)^{A_2}\left(\frac{x-x_r}{1-x_r}\right)^{A_3}
\left(\frac{x-x_\infty}{1-x_\infty}\right)^{2s+1}
f ^{\nu}_{\{0,1\};s}(x)\;, \nonumber\\
f ^{\nu}_{\{0,1\};s}(x)=\sum_{n=-\infty}^{\infty}
a_n^\nu F\left(-n-\nu+\frac{\omega}{2}-\frac{1}{2},
n+\nu+\frac{\omega}{2}+\frac{1}{2};\gamma;x\right)\;,
\ena
where $\nu$ is the characteristic exponent (the shifted angular 
momentum). Coefficients  $a_n^\nu$ are determined by solving 
the following three term recurrence relation with the initial condition 
 $a_0^\nu=1$,   
\bea
\alpha^\nu_n a^\nu_{n+1} + \beta^\nu_n a^\nu_n 
+ \gamma^\nu_n a^\nu_{n-1}=0, 
\ena
where
\bea
\alpha^\nu_n &=& -\frac{(n+\nu-\frac{\omega}{2}+\frac{3}{2})
(n+\nu-\sigma_+ +\frac{\omega}{2}+\frac{3}{2})
(n+\nu-\sigma_- +\frac{\omega}{2} +\frac{3}{2})
(n+\nu+\delta -\frac{\omega}{2}+\frac{1}{2})}
{2(n+\nu+1)(2n+2\nu+3)}, \nonumber\\
\beta^\nu_n &=&\frac{(1-\omega)(\gamma-\delta)
(\sigma_+ -\sigma_- +\epsilon-1)
(\sigma_+ -\sigma_- -\epsilon+1)}
{32(n+\nu)(n+\nu+1)}
+\left(\frac{1}{2}-x_r\right)(n+\nu)(n+\nu+1) \nonumber\\
&&     +\frac{1}{4}[\epsilon(\gamma-\delta)+\delta(1-\omega)
+2\sigma_+ \sigma_-]
+\frac{\omega^2-1}{4}x_r +v, \nonumber\\
\gamma^\nu_n &=& -\frac{(n+\nu+\sigma_+-\frac{\omega}{2} -\frac{1}{2})
(n+\nu+\sigma_- -\frac{\omega}{2} -\frac{1}{2})
(n+\nu+\gamma -\frac{\omega}{2}-\frac{1}{2}) 
(n+\nu+\frac{\omega}{2}-\frac{1}{2})} {2(n+\nu)(2n+2\nu-1)}.
\nonumber\\
\ena

\vskip 3mm  
\noindent
{\it (a-2) Determination of the characteristic exponent} 
 
The characteristic exponent $\nu$ is determined such that the 
series of hypergeometric functions converges.  
Let us  define the continued fractions
\begin{equation}
R_n(\nu)=\frac{a^\nu_n}{a^\nu_{n-1}}, \makebox[2cm]{}
L_n(\nu)=\frac{a^\nu_n}{a^\nu_{n+1}},
\end{equation}
which satisfy
\begin{equation}
R_n(\nu)=-\frac{\gamma^\nu_n}{\beta^\nu_n+\alpha^\nu_n R_{n+1}(\nu)}
 \;, \qquad
L_n(\nu) = -\frac{\alpha^\nu_n}{\beta^\nu_n+\gamma^\nu_n L_{n-1}(\nu)}
\;.
\end{equation}
Now we can evaluate the coefficients $a_n^{\nu}$ by using either 
series $R_n(\nu)$ or $L_n(\nu)$ with an appropriate initial data. 
The convergence of the series requires the 
following transcendental equation,
\begin{equation}
R_n(\nu) L_{n-1}(\nu) = 1. \label{eqn:aug}
\end{equation}
This is the equation to determine the characteristic exponent $\nu$. 
By using $\alpha_{-n}^{-\nu-1}=\gamma_n^\nu$ and 
$\beta_{-n}^{-\nu-1}=\beta_n^\nu$,  
we can prove that if $\nu$ is 
a solution of the  transcendental equation, then $-\nu-1$ is a 
solution, similarly to the proof given in Ref.[1]  for the Kerr 
geometry case. We can also prove that  
\bea
a_{-n}^{-\nu-1}=a_n^{\nu},
\ena
by using the recurrence relation and 
by taking the initial condition 
$a_0^{\nu}=a_0^{-\nu-1}$.  

\vskip 3mm 
\noindent
{\it (a-3) The solution which satisfies the outgoing boundary 
condition at the outer horizon}

This solution is simply given by 
\bea
 R^{\nu}_{out;\{0,1\};s}= (\Delta_r^{-s}  
 R^{\nu}_{in;\{0,1\};-s})^*\;,
\ena
which is proved in Appendix. The convergence region of this 
solution is the same as that for $R^{\nu}_{in;\{0,1\};s}$.

\vskip 3mm 
\noindent
{\it (a-4) Solutions specified by characteristic exponents}
 
By using the relation in Eq.(3.7), we can 
show that $R^{-\nu-1}_{in;\{0,1\};s}(x)=R^\nu_{in;\{0,1\};s}(x)$. 
This means 
that $R^\nu_{in;\{0,1\};s}(x)$ is expressed by a sum of two 
independent solutions with  characteristic exponents, 
$\nu$ and $-\nu-1$. We can explicitly show with $z=1-x$ that 
\bea
R^\nu_{in;\{0,1\};s}(x) =R^\nu_{\{0,1\};s} (z)
+R^{-\nu-1}_{\{0,1\};s} (z)\;,
\ena
where 
\bea
R^\nu_{\{0,1\};s} (z) &=& z^{A_2}(z-1)^{A_1}\left(1-\frac{z}{z_r}
\right)^{A_3}
\left(1-\frac{z}{z_\infty}\right)^{2s+1}
\Gamma(\gamma)
\nonumber\\
&&\times \sum_{n=-\infty}^{\infty} a_n^\nu
\frac{\Gamma(2n+2\nu+1)}
{\Gamma(n+\nu+\frac{\omega}{2}+\frac{1}{2})
\Gamma(n+\nu+\gamma-\frac{\omega}{2}+\frac{1}{2} )} 
z^{n+\nu-\frac{\omega}{2}+\frac{1}{2}}
 \nonumber\\
&&\times
F\left(-n-\nu+ \frac{\omega}{2}-\frac{1}{2},
 -n-\nu+\gamma- \frac{\omega}{2}-\frac{1}{2};
-2n-2\nu; \frac{1}{z}\right)\;. \nonumber\\
\ena 

\vskip 3mm 
\noindent
{\it (a-5) The convergence region of the series for 
$R^{\nu}_{in; \{0,1\};s}(x)$, etc.}

Firstly, we consider the convergence region for 
$R^{\nu}_{in; \{0,1\};s}(x)$. We find that 
\bea
\lim_{n\to \infty} \left|\frac{F_{n+1}^\nu}{F_{n}^\nu}\right|
=\lim_{n\to -\infty} \left|\frac{F_{n}^\nu}{F_{n+1}^\nu}\right|={\rm max} 
\{ |e^{\xi}|, |e^{-\xi}|\}\ge 1\;,
\ena
where we used the abbreviated expression $F_n^\nu$ for the 
hypergeometric function appeared in Eq.(3.1) as 
\bea
F_n^\nu \equiv F\left(-n-\nu+\frac{\omega}{2}-\frac{1}{2},
n+\nu+\frac{\omega}{2}+\frac{1}{2};\gamma;x\right)\;,
\ena
and 
\bea
e^{\pm\xi}=1-2x\pm\sqrt{(1-2x)^2-1}\;.
\ena
On the other hand, 
the ratio of coefficients converge 
either $e^{\xi_r}>1$ or $e^{-\xi_r}<1$. Taking account for 
Eq.(3.11), we have to require their limits as follows; 
\bea
\lim_{n\to \infty} \frac{a_{n+1}^\nu}{a_{n}^\nu}=
\lim_{n\to -\infty} \frac{a_{n}^\nu}{a_{n+1}^\nu}=e^{-\xi_r}\;,
\ena
where 
\bea
e^{\xi_r}=1-2x_r+\sqrt{(1-2x_r)^2-1}>1 \qquad (\; x_r<0\;)\;.
\ena
In order for the series of hypergeometric functions converges, 
it is required that 
\bea
{\rm max}\{|e^{\xi}|, |e^{-\xi}|\}<e^{\xi_r}\;.
\ena
 
Now we parametrize $x$ as $1-2x=(t+\frac1{t})/2$ where 
$|t|\ge 1$. This transformation maps a circle in the complex 
$t$ plane, $|t|=C$ ($C$:constant) to an ellipse in the 
complex $x$ plane.  
The condition of the convergence in Eq.(3.16) is expressed by 
$|t|<e^{\xi_r}$, i.e., inside the circle. In the complex 
$x$ plane, the convergence region is  inside 
the ellipse with foci 0 and 1, where the major axis is 
along  the real axis  extending from $x_r$ to $1-x_r$. 

The convergence region for $R^{\nu}_{out; \{0,1\};s}(x)$ is the 
same as $R^{\nu}_{in; \{0,1\};s}(x)$ as seen from Eq.(3.8). 
The convergence region for $R^\nu_{\{0,1\};s} (z)$ is also the 
same as $R^{\nu}_{in \{0,1\};s}(x)$, because 
$R^\nu_{\{0,1\};s} (z)$ and $R^{-\nu-1}_{\{0,1\};s} (z)$ are 
expressed by linear combinations of $R^{\nu}_{in;\{0,1\};s}$ and 
$R^{\nu}_{out;\{0,1\};s}$. 

In summary, the convergence region of series of all 
solutions is inside of the 
ellipse  with foci 0 and 1, where the major axis is along  
the real axis  extending from $x_r$ to $1-x_r$. 
If we confine $x$ to the physical region, the convergence region 
is $x>x_r$, i.e., $r<r_+'$,  with $r_+'$ being 
the de Sitter horizon.

\vskip 5mm 
\noindent
(B) Solutions  convergent around the de Sitter horizon 

Here, we use the variable $z$ rather than $x$ and construct 
solutions which converge around the de Sitter horizon. 

\vskip 3mm 
\noindent
{\it (b-1) Solutions which are specified by the characteristic 
exponents}

Firstly, we 
consider the solution with the characteristic exponent $\nu$. 
For this, we factor out the singularity at $z=z_{\infty}$ 
by the transformation 
\bea
R^{\nu}_{\{z_r,\infty\};s}(z)=
z^{A_2}(z-1)^{A_1}\left(1-\frac{z}{z_r}\right)^{A_3}
\left(1-\frac{z}{z_\infty}\right)^{2s+1} 
g^{\nu}_{\{z_r,\infty\};s}(z),
\ena
where $A_i \ (i=1,2,3)$ takes either $A_{i+}$  or $A_{i-}$. Then, 
$g^{\nu}_{\{z_r,\infty\};s}(z)$ should satisfy the following 
Heun's equation
\bea
\left\{ \ \frac{d^2}{dz^2} + \left[\frac{\delta}{z} +
\frac{\gamma}{z-1}
+\frac{\epsilon}{z-z_r} \right] \frac{d}{dz} 
+ \frac{\sigma_+ \sigma_- z -v-\sigma_+ \sigma_- }{z(z-1)(z-z_r)} 
\ \right\} g^{\nu}_{\{z_r,\infty\};s}(z)  =0.
\ena
Now we make the variable change $\zeta=z_r/z$ and the 
transformation 
$g^{\nu}_{\{z_r,\infty\};s}(z)=
\zeta^{\scriptstyle \sigma_+} h_{1;s}^\nu
(\zeta)$, we find a Heun's equation
\bea
&& \left\{ \ \frac{d^2}{d\zeta^2} 
+ \left[\frac{\sigma_+ -\sigma_- +1}{\zeta} 
+ \frac{\epsilon}{\zeta-1}
+\frac{\gamma}{\zeta-z_r} \right] \frac{d}{d\zeta} \right.\nonumber\\ 
&& \hspace{1cm} \left.
+ \frac{\sigma_+ (\sigma_+ +\gamma-\omega)\zeta 
- v-z_r \sigma_+ (\sigma_+ -\omega)-\sigma_+ \gamma}
{\zeta (\zeta-1)(\zeta-z_r)} \ \right\} h_{1;s}^{\nu}(\zeta)  =0\;,
\ena
where we used the Heun's constraint in Eq.(2.17). 
Then, we find  a solution 
\bea
g^{\nu}_{1;\{z_r,\infty\};s} &=&\frac
{\Gamma(\nu+\sigma_- -\frac{\omega}{2}+\frac{1}{2})}
{ \Gamma(\nu-\sigma_-+\frac{\omega}{2}+\frac{3}{2})} 
\left(\frac{z_r}{z}\right)^{\scriptstyle \sigma_+}
\sum_{n=-\infty}^{\infty} a_n^\nu
(-)^n \frac{ 
\Gamma(n+\nu-\sigma_-+\frac{\omega}{2}+\frac{3}{2})} 
{\Gamma(n+\nu+\sigma_- -\frac{\omega}{2}+\frac{1}{2})}
\nonumber\\
&&  \times F\left(-n-\nu+\sigma_+ -\frac{\omega}{2}-\frac{1}{2},
n+\nu+\sigma_+ -\frac{\omega}{2}+\frac{1}{2};
\sigma_+ -\sigma_- +1; \frac{z_r}{z}\right),
\nonumber\\
\ena
and the other by exchanging $\sigma_+$ with $\sigma_-$ 
\bea
g^{\nu}_{2;\{z_r,\infty\};s} =
g^{\nu}_{1;\{z_r,\infty\};s}\Big|_{\sigma_+ \iff \sigma_-}\;.
\ena
They satisfy the  relations 
$g^{-\nu-1}_{i;\{z_r,\infty\};s}=
g^{\nu}_{i;\{z_r,\infty\};s}$ \ (i=1, 2), 
so that a combination of them forms a solution with the 
characteristic exponent $\nu$. In fact, we find    
\bea
g^{\nu}_{\{z_r,\infty\};s}&=&
\frac{(-)^{-\nu +\frac{\omega}{2}-\frac{1}{2}}
}{\Gamma(\sigma_+ -\epsilon+1)}
\left[(-)^{ -\sigma_+  }\frac{\Gamma(\sigma_- -\sigma_+)
\Gamma(\nu-\sigma_-+\frac{\omega}{2}+\frac32)}
{\Gamma(\nu+\sigma_--\frac{\omega}{2}+\frac12)}
g^{\nu}_{1;\{z_r,\infty\};s} \right.
\nonumber\\
&& \hskip 1cm
\left.+(-)^{-\sigma_- }\frac{\Gamma(\sigma_+ -\sigma_-)
\Gamma(\nu-\sigma_+ +\frac{\omega}{2}+\frac32)}
{\Gamma(\nu+\sigma_+ -\frac{\omega}{2}+\frac12)}
g^{\nu}_{2;\{z_r,\infty\};s} \right]\;.
\nonumber\\
\ena
gives the solution.  By substituting this, we find 
\bea
R^{\nu}_{\{z_r,\infty\};s}(z)
&=&z^{A_2}(z-1)^{A_1}\left(1-\frac{z}{z_r}\right)^{A_3}
\left(1-\frac{z}{z_\infty}\right)^{2s+1} 
\frac{1}{\Gamma(\sigma_+ -\epsilon+1)}
\nonumber\\
&& \makebox[-10mm]{}\times
\sum_{n=-\infty}^{\infty} a_n^\nu
\frac{\Gamma(n+\nu-\sigma_++\frac{\omega}{2}+\frac{3}{2})
\Gamma(n+\nu-\sigma_- +\frac{\omega}{2}+\frac{3}{2})}
{\Gamma(2n+2\nu+2)} \left(\frac{z}{z_r}\right)^{n+\nu-\frac{\omega}{2}
+\frac{1}{2}}
\nonumber\\
&&  \makebox[-10mm]{}
\times  F\left(n+\nu+\sigma_+ -\frac{\omega}{2}+\frac{1}{2},
n+\nu+\sigma_- -\frac{\omega}{2}+\frac{1}{2};
2n+2\nu+2;
\frac{z}{z_r}\right).\nonumber\\
\ena
is a solution specified by the characteristic exponent $\nu$. 
Another independent solution with the characteristic 
exponent $-\nu-1$ is given by 
\bea
R^{-\nu-1}_{\{z_r,\infty\};s}(z).
\ena

{\it (b-2) Solutions which satisfy 
the incoming and the outgoing boundary conditions 
at the de Sitter horizon}

The solution $R^{\nu}_{\{z_r,\infty\};s}(z)$ is 
expressed by the sum of solutions which satisfy 
the incoming and the outgoing boundary conditions 
at the de Sitter horizon. We find 
\bea
R^{\nu}_{\{z_r,\infty\};s}(z)&=&
R^{1;\nu}_{\{z_r,\infty\};s}(z)
+R^{2;\nu}_{\{z_r,\infty\};s}(z), 
\ena
where 
\bea
R^{1;\nu}_{\{z_r,\infty\};s}(z) &=&
z^{A_2}(z-1)^{A_1}\left(1-\frac{z}{z_r}\right)^{A_3-\epsilon+1}
\left(1-\frac{z}{z_\infty}\right)^{2s+1} 
\frac{\Gamma(\epsilon-1)}{\Gamma(\sigma_+ -\epsilon+1)} 
\nonumber\\
&& \makebox[-10mm]{} \times\sum_{n=-\infty}^{\infty}a_n^\nu
\frac{\Gamma(n+\nu-\sigma_+ +\frac{\omega}{2}+\frac{3}{2})
\Gamma(n+\nu-\sigma_- +\frac{\omega}{2}+\frac{3}{2})
} {\Gamma(n+\nu+\sigma_+ -\frac{\omega}{2}+\frac12)
\Gamma(n+\nu+\sigma_- -\frac{\omega}{2}+\frac{1}{2})} 
\left(\frac{z}{z_r}\right)^{n+\nu-\frac{\omega}{2}+\frac{1}{2}}
\nonumber\\
&& \makebox[-10mm]{} \times
F\left(n+\nu-\sigma_+ +\frac{\omega}{2}+\frac{3}{2},
n+\nu-\sigma_- +\frac{\omega}{2}+\frac{3}{2};
2-\epsilon;
1-\frac{z}{z_r} \right), 
\nonumber\\
\ena
and 
\bea
R^{2;\nu}_{\{z_r,\infty\};s}(z) &=&
z^{A_2}(z-1)^{A_1}\left(1-\frac{z}{z_r}\right)^{A_3}
\left(1-\frac{z}{z_\infty}\right)^{2s+1} 
\frac{\Gamma(1-\epsilon)}{\Gamma(\sigma_+ -\epsilon+1)}
\sum_{n=-\infty}^{\infty} a_n^\nu
\nonumber\\
&& \makebox[-13mm]{} \times 
\left(\frac{z}{z_r}\right)^{n+\nu-\frac{\omega}{2}+\frac{1}{2}}
F\left(n+\nu+\sigma_+ -\frac{\omega}{2}+\frac{1}{2},
n+\nu+\sigma_- -\frac{\omega}{2}+\frac{1}{2};
\epsilon;
1-\frac{z}{z_r} \right).
\nonumber\\
\ena

\vskip 3mm 
\noindent
{\it (b-3) The convergence region for 
$R^{\nu}_{\{z_r,\infty\};s}(z)$, 
etc. }
 
Firstly, we consider $g^\nu_{i;\{z_r,\infty \};s}$ $(i=1,2)$ in 
Eqs.(3.20) and (3.21). By 
comparing  Eqs.(3.20) and (3.21) with  (3.1), we find that 
the convergence region of the series for 
$g^\nu_{i;\{z_r,\infty\};s}(z)$ $(i=1,2)$ in complex $z_r/z$ plane is 
inside the ellipse with foci are 0 and 1, where with the major axis 
is along 
the real axis  extending from $x_r$ to $1-x_r$, 
i.e., $x_r<z_r/z<1-x_r$.  Keeping in mind that $z=1-x$ 
and $z_r=1-x_r$, we find that the convergence region is $x<0$, 
if we confine $x$ 
to physical values. This region corresponds to  $r>r_+$ with 
$r_+$ being the outer horizon.  

Now we observe that $R^{\nu}_{\{z_r,\infty\};s}(z)$ and 
$R^{-\nu-1}_{\{z_r,\infty\};s}(z)$ are expressed by linear 
combination of $g^\nu_{i;\{z_r,\infty\};s}(z)$ $(i=1,2)$, and also 
$R^{1;\nu}_{\{z_r,\infty\};s}(z)$ and
$R^{2;\nu}_{\{z_r,\infty\};s}(z)$ 
are expressed by linear combinations of 
$R^{\nu}_{\{z_r,\infty\};s}(z)$ and
$R^{-\nu-1}_{\{z_r,\infty\};s}(z)$. 
Thus, we conclude that the convergence region of all these functions 
are the same region, i.e., $r>r_+$.  

\vskip 3mm
\noindent
(C) The matching of two solutions

We consider the matching of  the solution which is convergent 
for $r<r_+'$ 
and the solution which is convergent for $r>r_+$.  We make the 
matching in the region $r_+<r<r_+'$ where  
both series are convergent.  The matching 
can be made between two solutions with the 
same characteristic exponent. Explicitly, we require that 
\bea
R^\nu_{\{0,1\};s} (z)=K_\nu R^{\nu}_{\{z_r,\infty\};s}(z).
\ena
for the region $r_+<r<r_+'$. 
The proportionality constant 
$K_\nu$ is determined by comparing the coefficients of 
 $z^{p+\nu-\frac{\omega}{2}+\frac{1}{2}}$. The result is 
\bea
K_\nu \makebox[-3mm]{}&=&\makebox[-3mm]{}
\frac{z_r^{p+\nu-\frac{\omega}{2}+\frac{1}{2}}
\Gamma(\gamma)\Gamma(\sigma_+ -\epsilon+1)} 
{\Gamma(p+\nu-\frac{\omega}{2}+\frac{3}{2}) 
\Gamma(p+\nu+\delta-\frac{\omega}{2}+\frac{1}{2})
\Gamma(p+\nu+\sigma_+-\frac{\omega}{2}+\frac{1}{2})
\Gamma(p+\nu+\sigma_--\frac{\omega}{2}+\frac{1}{2})} 
\nonumber\\
&& \makebox[-10mm]{}\times
\left[\sum_{n=p}^{\infty} a_n^\nu
\frac{(-)^{n-p}\Gamma(n+\nu-\frac{\omega}{2}+\frac{3}{2})
\Gamma(n+\nu+\delta-\frac{\omega}{2}+\frac{1}{2})
\Gamma(p+n+2\nu+1)}
{\Gamma(n+\nu+\frac{\omega}{2}+\frac{1}{2})
\Gamma(n+\nu+\gamma-\frac{\omega}{2}+\frac{1}{2})(n-p)!} \right] 
\nonumber\\
&&\makebox[-10mm]{} \times
\left[\sum_{n=-\infty}^{p} a_n^\nu
\frac{\Gamma(n+\nu-\sigma_+ +\frac{\omega}{2}+\frac{3}{2})
\Gamma(n+\nu-\sigma_- +\frac{\omega}{2}+\frac{3}{2})} 
{\Gamma(n+\nu+\sigma_+ -\frac{\omega}{2}+\frac{1}{2}) 
\Gamma(n+\nu+\sigma_- -\frac{\omega}{2}+\frac{1}{2})
\Gamma(p+n+2\nu+2)(p-n)!} \right]^{-1}.
\nonumber\\
\ena
The constant $K_\nu$ includes an integer $p$, but  is 
independent of it. 

\vskip 3mm
\noindent
(D) Solutions valid in the entire region

The solution which satisfies the incoming boundary 
condition at the outer horizon of black hole and 
convergent in the entire region is expressed by
\bea
R^{\nu}_{in;s}&=&\tilde A_s R^{\nu}_{in;\{0,1\};s} \nonumber \\ 
&=&\tilde A_s \left[K_{\nu}(s) R^{\nu}_{\{z_r,\infty\};s}(z)
+K_{-\nu-1}(s) R^{-\nu-1}_{\{z_r,\infty\};s}(z)\right],
\ena
where $\tilde A_s$ is the normalization constant. 
The first can be  used  for $r<r'_+$ with $r_+'$ being the de Sitter 
horizon    and the second 
can be  used for $r>r_+$ with $r_+$ being the outer horizon. 

The other solution which satisfies the other boundary condition 
is obtained from a linear combination of two independent 
solutions, $R^{\nu}_{in;s} $ and 
\bea
 R^{\nu}_{out;s}= (\Delta_r^{-s}  R^{\nu}_{in;-s})^*\;.
\ena
which satisfies the outgoing boundary condition at the outer 
horizon. 
\section{ The Teukolsky-Starobinsky identities}
\setcounter{equation}{0} 

We discuss the Teukolsky-Starobinsky (T-S) identities for 
perturbed fields in Kerr-Newman-de Sitter geometries, 
based on Teukolsky equation defined in Eq.(2.3), although 
this equation is not applied for electromagnetic 
field and gravitational field because they couple each other 
in this geometry. Nevertheless, we make the analysis 
because we want to see the mathematical structure of 
Teukolsky equation in detail and also we want to treat 
the T-S identities in the unified manner. 
Of course, the following analysis is  valid for spin 1 
fields which do not couple to electric charge and 
needless to say that it is valid for all fields 
for Kerr-de Sitter geometries. The T-S identities are 
expressed by 
\bea
\Delta_r^s({\cal D}^\dagger_Q)^{2s}\Delta_r^s R_s &=& C_s^*R_{-s}
\qquad \mbox{(T-S identity (A))}, 
\nonumber\\ 
({\cal D}_Q)^{2s} R_{-s} 
&=& C_sR_{s}
\qquad \mbox{(T-S identity (B))}\;,
\ena
where $C_s$ are Starobinsky constants and 
${\cal D}_Q$ and ${\cal D}^\dagger_Q$ are defined by 
\bea
{\cal D}^\dagger_Q&=&\partial_r+i\frac{1+\alpha}{\Delta_r}
\left(K-\frac{eQ}{1+\alpha}r\right)\;,\nonumber\\
{\cal D}_Q&=&\partial_r-i\frac{1+\alpha}{\Delta_r}
\left(K-\frac{eQ}{1+\alpha}r\right)\;.
\ena

The T-S identities are interesting because of their 
mathematical structure as well as to fix the relative 
normalization between the solutions specified by $s$ and 
$-s$.

\vskip 3mm
\noindent
(A) Differential operators

Since the solutions are expressed in terms of $x$ or $z$, 
we rewrite the differential operator ${\cal D}_Q^\dagger$ as
\bea
{\cal D}^\dagger_Q 
&=& -\frac{r_+ -r'_-}{(r_+ -r_-)(r_- -r'_-)}
(-x)^{-a_1}(1-x)^{-a_2}
\left(\frac{x-x_r}{1-x_r}\right)^{-a_3}
\left(\frac{x-x_\infty}{1-x_\infty}\right)^2 \nonumber\\
&&  \times \frac{d}{dx}(-x)^{a_1}(1-x)^{a_2}
\left(\frac{x-x_r}{1-x_r}\right)^{a_3}\nonumber\\
&=& \frac{r_+ -r'_-}{(r_+ -r_-)(r_- -r'_-)}
z^{-a_2}(z-1)^{-a_1}
\left(1-\frac{z}{z_r}\right)^{-a_3}
\left(1-\frac{z}{z_\infty}\right)^2 \nonumber \\
&&  \times \frac{d}{dz}z^{a_2}(z-1)^{a_1}
\left(1-\frac{z}{z_r}\right)^{a_3}.
\ena
The ${\cal D}_Q$  is derived by taking the complex 
conjugation of ${\cal D}^\dagger_Q$.

\vskip 3mm
\noindent
(B) Parametrizations of solutions

Solutions are specified by the choice of $A_i$ in Eq.(2.11) 
so that we have to specify these parameters to fix them. 

\vskip 3mm
\noindent
{\it (b-1) A specific choice of $A_{i}$}

Here we consider the solution which satisfies the incoming 
boundary condition at the outer horizon and thus we choose 
$A_1=A_{1-}$. For others, we choose $A_2=A_{2+}$ and 
$A_3=A_{3-}$. Then we have 
\bea 
A_1=-a_1-s,\; A_2=a_2,\;A_3=-a_3-s\;, 
\ena
so that we find
\bea
\sigma_+&=&2a_2-s+1, \;\sigma_-=-2a_1-2a_3+1,\nonumber\\
\gamma&=&-2a_1-s+1, \;\delta=2a_2+s+1, \;\epsilon=-2a_3-s+1\;, 
\nonumber\\
\omega&\equiv& \gamma+\delta-1=\sigma_+ +\sigma_- -\epsilon
=-2a_1+2a_2+1\;.
\ena

Then, from Eqs.(3.1) and (3.30) the solution  is given by the 
following two expressions which have  different regions of 
convergence,
\bea
R^{\nu}_{in;s}&=&\tilde A_s(-x)^{-s-a_1}(1-x)^{a_2}
\left(\frac{x-x_r}{1-x_r}\right)^{-s-a_3}
\left(\frac{x-x_\infty}{1-x_\infty}\right)^{2s+1} 
\nonumber\\ && \times 
\sum_{n=-\infty}^{\infty}a^\nu_n(s)
F(-n-\nu-a_1+a_2,n+\nu-a_1+a_2+1;-2a_1-s+1;x) \;,
\nonumber\\
\ena
which is convergent for $r<r'_+$ (the expression (A)) and
\bea
R^{\nu}_{in;s} = \tilde A_s 
\left[K_{\nu}(s) R^{\nu}_{\{z_r,\infty\};s}(z)
+K_{-\nu-1}(s) R^{-\nu-1}_{\{z_r,\infty\};s}(z)\right]\;,
\nonumber\\
\ena
where
\bea
R^{\nu}_{\{z_r,\infty\};s}(z)&&\makebox[-7mm]{}
=z^{a_2}(z-1)^{-s-a_1}
\left(1-\frac{z}{z_r}\right)^{-s-a_3}
\left(1-\frac{z}{z_\infty}\right)^{2s+1} \nonumber \\
&& \makebox[-20mm]{}
\times \sum_{n=-\infty}^{\infty}a^\nu_n(s)
\frac{\Gamma(n+\nu+a_3-a_4+1)
\Gamma(n+\nu-a_1-a_2+s+1)} {\Gamma(2a_2+2a_3+1)
\Gamma(2n+2\nu+2)}
\left(\frac{z}{z_r}\right)^{n+\nu+a_1-a_2} \nonumber \\
&& \makebox[-20mm]{}
\times F\left(n+\nu+a_1+a_2-s+1,n+\nu-a_3+a_4+1;
2n+2\nu+2;\frac{z}{z_r}\right)\;,
\nonumber\\
\ena
which is convergent for $r>r_+$ (the expression (B)).

\vskip 3mm
\noindent
{\it (b-2) Another  choice of $A_{i}$}

It may be interesting to ask the relation between  solutions 
with different choices of $A_i$. To answer this question, we 
consider a generic Heun's equation,
\bea
\left\{\frac{d^2}{dz^2}+\left[\frac{\gamma}{z}
+\frac{\delta}{z-1}+\frac{\epsilon}{z-a_H}\right]\frac{d}{dz}
+\frac{\sigma_+ \sigma_- z-q} {z(z-1)(z-a_H)}\right\}f(z)=0, 
\ena 
with $\sigma_+ +\sigma_- -\epsilon=\gamma+\delta-1$.  Here we 
express a particular solution as 
$Hf(a_H, q; \sigma_+, \sigma_-, \gamma, \delta; z)$ 
which is regular at $z=0$ and is normalized 
to be one at $z=0$. 
By setting 
\bea
f(z)=\left(1-\frac{z}{a_H}\right)^{1-\epsilon}g(z)\;,
\ena
the above Heun's equation is transformed to another form as
\bea
&&\left\{\frac{d^2}{dz^2}+\left[\frac{\gamma}{z}
+\frac{\delta}{z-1}+\frac{2-\epsilon}{z-a_H}\right]\frac{d}{dz}
\right. \nonumber\\
&& \hskip 30mm 
\left. +\frac{(\gamma+\delta-\sigma_+)(\gamma+\delta-\sigma_-)z
- q-\gamma(1-\epsilon)}{z(z-1)(z-a_H)}\right\}g(z)=0\;,
\nonumber\\
\ena
where we used $1-\epsilon=\gamma+\delta-\sigma_+-\sigma_-$. 
The solution which is normalized and regular at $z=0$ is given by 
$Hf(a_H,q+\gamma(1-\epsilon);
\gamma+\delta-\sigma_+,\gamma+\delta-\sigma_-, \gamma,\delta;z)$. 
Now we have two different expressions of the solution of Eq.(4.9). 
Both are 
regular at $z=0$ and are normalized to be one at $z=0$, so that two 
solutions should be equal each other, 
\bea
&&Hf(a, q; \sigma_+, \sigma_-, \gamma, \delta; z)
\nonumber\\
&&\hskip 10mm  =\left(1-\frac{z}{a_H}\right)^{1-\epsilon}
Hf(a_H,q+\gamma(1-\epsilon);\gamma+\delta-\sigma_+,
\gamma+\delta-\sigma_-,\gamma,\delta;z)\;.
\nonumber\\
\ena

We apply this fact to the incoming solution at outer horizon in 
Eq.(4.6), which  is expressed by
\bea
 R^{\nu}_{in;s} &=& \left(\tilde A_s 
\sum_{n=-\infty}^{\infty}a^{\nu}_n(s)\right)
(-x)^{-s-a_1}(1-x)^{a_2}
\left(\frac{x-x_r}{1-x_r}\right)^{-s-a_3}
\left(\frac{x-x_\infty}{1-x_\infty}\right)^{2s+1} \nonumber \\
&& \times Hf(x_r,-v;\sigma_+,\sigma_-,\gamma,\delta;x),
\ena
where
\bea
&&Hf(x_r,-v;\sigma_+,\sigma_-,\gamma,\delta;x) = \nonumber \\
&&\hskip 10mm  \left[\sum_{n=-\infty}^{\infty}a^{\nu}_n(s)\right]^{-1}
\sum_{n=-\infty}^{\infty} a^{\nu}_n(s)
F(-n-\nu+\frac{\omega}{2}-\frac{1}{2},n+\nu+\frac{\omega}{2}+\frac{1}{2};
\gamma;x)\;,
\nonumber\\
\ena
By using the relation between  Heun's functions in Eq.(4.12), 
this solution is at the same time expressed by 
\bea
R^{\nu}_{in;s} &=& \tilde A_s
\left(\frac{-x_r}{1-x_r}\right)^{-s-2a_3}
\left[\sum_{n=-\infty}^{\infty}a^{\nu}_n(s)\right]
(-x)^{-s-a_1}(1-x)^{a_2}
\left(\frac{x-x_r}{1-x_r}\right)^{a_3}
\left(\frac{x-x_\infty}{1-x_\infty}\right)^{2s+1} \nonumber \\
&& \times Hf(a_H,-v'+\gamma(1-\epsilon);\gamma+\delta-\sigma_+,
\gamma+\delta-\sigma_-,\gamma,\delta;x)\;, 
\ena
where
\bea
&& Hf(a_H,-v'+\gamma(1-\epsilon);\gamma+\delta-\sigma_+,
\gamma+\delta-\sigma_-,\gamma,\delta;x)\nonumber \\
&& \hskip 1cm
=\left[\sum_{n=-\infty}^{\infty}b^{\nu}_n(s)\right]^{-1} 
\sum_{n=-\infty}^{\infty}b^{\nu}_n(s)
F(-n-\nu+\frac{\omega}{2}-\frac{1}{2},
n+\nu+\frac{\omega}{2}+\frac{1}{2};\gamma;x)\;.
\nonumber\\
\ena
Here the coefficients are defined by
\bea
b^{\nu}_n(s) &=& a^{\nu}_n(s)
\frac{\Gamma(\nu+\sigma_+ -\frac{\omega}{2}+\frac{1}{2})
\Gamma(\nu+\sigma_- -\frac{\omega}{2}+\frac{1}{2})} 
{\Gamma(\nu+\frac{\omega}{2}-\sigma_+ +\frac{3}{2})
\Gamma(\nu+\frac{\omega}{2}-\sigma_- +\frac{3}{2})} \nonumber \\
&& \times
\frac{\Gamma(n+\nu+\frac{\omega}{2}-\sigma_+ +\frac{3}{2})
\Gamma(n+\nu+\frac{\omega}{2}-\sigma_- +\frac{3}{2})}
 {\Gamma(n+\nu+\sigma_+ -\frac{\omega}{2}+\frac{1}{2})
\Gamma(n+\nu+\sigma_- -\frac{\omega}{2}+\frac{1}{2})}\;.
\ena
The difference of coefficients are due to 
the change of parameters in the recurrence relation which  
coefficients satisfy. This solution in Eq.(4.16) 
is the one which corresponds to  
$A_1=A_{1-}$, $A_2=A_{2+}$ and $A_3=A_{3+}$.

\vskip 3mm
\noindent
(C) Various relations

The characteristic exponent $\nu$, coefficients $a_n^\nu$,  
$b_n^\nu$ and the proportionality constant $K_\nu$ satisfy 
the following relations, which are important to show that 
the solutions satisfy the T-S identities. 

\vskip 2mm
\noindent
{\it (c-1) The relation between $\nu(s)$ and $\nu(-s)$}

As we prove in Appendix, we have   
\bea
\nu(s)=\nu(-s)\;, \; \nu(s)=\nu(s)^*\;.
\ena 
These relation is vital for the T-S identities, because the 
T-S identities are differential transformation from the solution 
specified by $s$ to that by $-s$.  

\vskip 2mm
\noindent
{\it (c-2) The relation between coefficients and $K_\nu$}

By examining the recurrence relation in Eq.(3.2), we find 
\bea
a^\nu_n(-s)=
\left| \frac{\Gamma(\nu+a_1+a_2-s+1)}
{\Gamma(\nu+a_1+a_2+s+1)} \right|^2
\left| \frac{\Gamma(n+\nu+a_1+a_2+s+1)}
{\Gamma(n+\nu+a_1+a_2-s+1)} \right|^2 a^\nu_n(s)\;, 
\ena
where we chose $a^\nu_0(-s)=a^\nu_0(s)$. 
Also, by using this  relation, we find 
\bea
b^{\nu}_n(-s)=b^{\nu}_n(s)\;.
\ena 
and
\bea
\frac{K_{\nu}(s)}{K_{\nu}(-s)}=
\frac{K_{-\nu-1}(s)}{K_{-\nu-1}(-s)}
=\frac{\Gamma(-2a_1-s+1)}{\Gamma(-2a_1+s+1)}\;.
\ena

\vskip 2mm
\noindent
{\it (c-3) Useful mathematical formula}

The following mathematical relations are needed:
\bea
\left[\left(\frac{x-x_\infty}{1-x_\infty}\right)^2
\frac{d}{dx}\right]^k
= \left(\frac{x-x_\infty}{1-x_\infty}\right)^{k+1}
\left(\frac{d}{dx}\right)^k
\left(\frac{x-x_\infty}{1-x_\infty}\right)^{k-1}\;,
\nonumber
\ena 
\bea
&&\left(\frac{d}{dx}\right)^k(1-x)^{A+B-C}F(A,B;C;x)
\nonumber\\
&&\hskip 1cm =\frac{\Gamma(C-A+k)\Gamma(C-B+k)\Gamma(C)}
{\Gamma(C-A)\Gamma(C-B)\Gamma(C+k)}  
  (1-x)^{A+B-C-k}F(A,B;C+k;x)\;,
  \nonumber
\ena
\bea
\left(\frac{d}{dz}\right)^k 
z^{A+k-1}F(A,B;C;z)
=\frac{\Gamma(A+k)}{\Gamma(A)}
z^{A-1}F(A+k,B;C;z)\;.
\nonumber\\
\ena

\vskip 3mm
\noindent
(D) Teukolsky-Starobinsky identity 

First we discuss the solution given by the expression (A) in 
Eq.(4.6) and show analytically that this solution satisfies 
the T-S identity (A) when we take  the relative normalization 
between $R^{\nu}_{in;s}$ and $R^{\nu}_{in;-s}$ properly. 
In other words, we can fix  the normalization 
factor for $R^{\nu}_{in;s}$, $\tilde A_s$ for $s>0$ by the 
T-S identity once we fix $\tilde A_{-s}=1$. First we apply 
the operator $\Delta_r^s ({\cal D}^\dagger_Q)^{2s}\Delta_r^s$ 
to the expression (A):
\bea
\Delta_r^s && \makebox[-10mm]{}({\cal D}^\dagger_Q)^{2s}\Delta_r^s
R^{\nu}_{in;s} \nonumber \\ 
\makebox[-3mm]{}&=&\makebox[-3mm]{}
\tilde A_s \left[-\frac{\alpha}{a^2}(r_+ -r_-)(r'_+ -r_- )
(r_- -r'_-)\right]^{2s}
(-x)^{s-a_1}(1-x)^{s-a_2}
\left(\frac{x-x_r}{1-x_r}\right)^{s-a_3} \nonumber \\
&& \times   \left(\frac{x-x_\infty}{1-x_\infty}\right)^{-4s}
\left[\left(\frac{x-x_\infty}{1-x_\infty}\right)^2
\frac{d}{dx}\right]^{2s}
\left(\frac{x-x_\infty}
{1-x_\infty}\right)^{-2s+1}(1-x)^{s+2a_2} \nonumber\\ && \times 
\sum_{n=-\infty}^{\infty}a^\nu_n(s)
F(-n-\nu-a_1+a_2,n+\nu-a_1+a_2+1;-2a_1-s+1;x) \nonumber \\
\makebox[-3mm]{}&=&\makebox[-3mm]{}
\tilde A_s \left[-\frac{\alpha}{a^2}(r_+ -r_-)(r'_+ -r_- )
(r_- -r'_-)\right]^{2s} \nonumber \\
&& \times  (-x)^{s-a_1}(1-x)^{s-a_2}
\left(\frac{x-x_r}{1-x_r}\right)^{s-a_3}
\left(\frac{x-x_\infty}{1-x_\infty}\right)^{-2s+1}
\left(\frac{d}{dx}\right)^{2s}(1-x)^{s+2a_2} \nonumber \\
 && \times   \sum_{n=-\infty}^{\infty}a^\nu_n(s)
               F(-n-\nu-a_1+a_2,n+\nu-a_1+a_2+1;-2a_1-s+1;x) \nonumber
\\
\makebox[-3mm]{}&=&\makebox[-3mm]{}
\tilde A_s \left[\frac{\alpha}{a^2}(r_+ -r_-)(r'_+ -r_- )
(r_- -r'_-)\right]^{2s} \nonumber \\
&& \times
\frac{\Gamma(-2a_1-s+1)}{\Gamma(-2a_1+s+1)}
\left| \frac{\Gamma(\nu+a_1+a_2+s+1)}
{\Gamma(\nu+a_1+a_2-s+1)} \right|^2
R^{\nu}_{in;-s},
\ena
where we used  relations in Eqs.(4.22) and (4.19). 
Thus $\tilde A_s$ is fixed by 
\bea
\tilde A_s &=& C_s^* \left[
-\frac{a^2}{\alpha(r_+ -r_-)(r'_+ -r_-)(r'_- -r_-)}\right]^{2s}
\nonumber \\
&& \times \frac{\Gamma(-2a_1+s+1)}{\Gamma(-2a_1-s+1)}
\left| \frac{\Gamma(\nu+a_1+a_2-s+1)}
{\Gamma(\nu+a_1+a_2+s+1)} \right|^2\;(s>0)\;,
\ena
provided $\tilde A_{-s}=1\; (s>0)$.

Next we show that the expression (B) of the solution in Eq.(4.7) 
satisfies the T-S identity (A). 
To the end we observe for $s>0$
\bea
\Delta_r^s && \makebox[-10mm]{}({\cal D}^\dagger_Q)^{2s}\Delta_r^s
R^{\nu}_{\{z_r,\infty\};s}\nonumber \\ 
\makebox[-3mm]{}&=&\makebox[-3mm]{}
\left[\frac{\alpha}{a^2}(r_+ -r_-)( r'_+ -r_-)
(r_- -r'_-)\right]^{2s}
z^{s-a_2}(z-1)^{s-a_1}\left(1-\frac{z}{z_r}\right)^{s-a_3} \nonumber
\\
&& \times   \left(1-\frac{z}{z_\infty}\right)^{-4s}
\left[\left(1-\frac{z}{z_\infty}\right)^2
\frac{d}{dz}\right]^{2s}
\left(1-\frac{z}{z_\infty}\right)^{-2s+1}z^{s+2a_2} \nonumber \\ 
&& \times 
\sum_{n=-\infty}^{\infty}a^\nu_n(s)
\frac{\Gamma(n+\nu+a_3-a_4+1)\Gamma(n+\nu-a_1-a_2+s+1)}
{\Gamma(2a_2+2a_3+1)\Gamma(2n+2\nu+2)}\nonumber \\
&& \times  \left(\frac{z}{z_r}\right)^{n+\nu+a_1-a_2}
F\left(n+\nu+a_1+a_2-s+1,n+\nu-a_3+a_4+1;
2n+2\nu+2;\frac{z}{z_r}\right) \nonumber \\
\makebox[-3mm]{}&=&\makebox[-3mm]{}
\left[\frac{\alpha}{a^2}(r_+ -r_-)(r'_+ -r_- )
(r_- -r'_-)\right]^{2s} \nonumber \\
&& \times  z^{s-a_2}(z-1)^{s-a_1}
\left(1-\frac{z}{z_r}\right)^{s-a_3}
\left(1-\frac{z}{z_\infty}\right)^{-2s+1} \nonumber \\
&& \times \left(\frac{d}{dz}\right)^{2s}z^{s+2a_2}
\sum_{n=-\infty}^{\infty}a^\nu_n(s)
\frac{\Gamma(n+\nu+a_3-a_4+1)\Gamma(n+\nu-a_1-a_2+s+1)}
{\Gamma(2a_2+2a_3+1)\Gamma(2n+2\nu+2)} \nonumber \\
&& \times  \left(\frac{z}{z_r}\right)^{n+\nu+a_1-a_2}
F\left(n+\nu+a_1+a_2-s+1,n+\nu-a_3+a_4+1;
2n+2\nu+2;\frac{z}{z_r}\right) \nonumber \\
\makebox[-3mm]{}&=&\makebox[-3mm]{}
\left[\frac{\alpha}{a^2}(r_+ -r_-)(r'_+-r_- )
(r_- -r'_-)\right]^{2s} \left| \frac{\Gamma(\nu+a_1+a_2+s+1)}
{\Gamma(\nu+a_1+a_2-s+1)} \right|^2
R^{\nu}_{\{z_r,\infty\};-s}\;,
\nonumber\\
\ena
where we used relations in Eqs.(4.22) and (4.19).  

By exchanging $\nu$ with $-\nu-1$ and using the fact that $2s$ is 
integer, we find from Eq.(4.25) that 
\bea
&&\Delta_r^s ({\cal D}^\dagger_Q)^{2s}\Delta_r^s
R^{-\nu-1}_{\{z_r,\infty\};s} \nonumber\\
&&\hskip 1cm =  
\left[\frac{\alpha}{a^2}(r_+ -r_-)(r'_+ -r_- )
(r_- -r'_-)\right]^{2s} \left|\frac{\Gamma(\nu+a_1+a_2+s+1)}
{\Gamma(\nu+a_1+a_2-s+1)} \right|^2 R^{-\nu-1}_{\{z_r,\infty\};-s}\;.
\nonumber\\
\ena
Thus we find
\bea
\Delta_r^s ({\cal D}^\dagger_Q)^{2s}\Delta_r^s
R^{\nu}_{in;s} &=&\tilde A_s 
\left[\frac{\alpha}{a^2}(r_+ -r_-)(r'_+ -r_- )(r_- -r'_-)\right]^{2s}
\nonumber \\
&& \times
\frac{\Gamma(-2a_1-s+1)}{\Gamma(-2a_1+s+1)}
\left| \frac{\Gamma(\nu+a_1+a_2+s+1)}
{\Gamma(\nu+a_1+a_2-s+1)} \right|^2 \nonumber  \\ && \times 
\left[K_{\nu}(-s) R^{\nu}_{\{z_r,\infty\};-s}(z)
+K_{-\nu-1}(-s) R^{-\nu-1}_{\{z_r,\infty\};-s}(z)\right],
\nonumber \\
&=& C_s^* \left[K_{\nu}(-s) R^{\nu}_{\{z_r,\infty\};-s}(z)
+K_{-\nu-1}(-s) R^{-\nu-1}_{\{z_r,\infty\};-s}(z)\right],
\nonumber \\
&=& C_s^* R^{\nu}_{in;-s}(z),
\ena
where we used the relation in Eqs.(4.21). 

Next we examine the T-S identity (B) by using another 
parametrization of the solution given in Eq.(4.15). 
By applying the differential operator $ ({\cal D}_Q)^{2s}$ 
to this expression, we find 
\bea
({\cal D}_0)^{2s}R^{\nu}_{in;-s}\makebox[-3mm]{}&=&\makebox[-3mm]{}
\left(\frac{-x_r}{1-x_r}\right)^{s-2a_3}
\left[\sum_{n=-\infty}^{\infty}a^{\nu}_n(-s)\right]
\left[\sum_{n=-\infty}^{\infty}b^{\nu}_n(-s)\right]^{-1}
\left[-\frac{r_+ -r'_-}{(r_+ -r_-)(r_- -r'_-)}\right]^{2s} \nonumber
\\
&& \times (-x)^{a_1}(1-x)^{a_2}\left(\frac{x-x_r}{1-x_r}\right)^{a_3}
\left[
\left(\frac{x-x_\infty}{1-x_\infty}\right)^{2}\frac{d}{dx}\right]^{2s}
\left(\frac{x-x_\infty}{1-x_\infty}\right)^{-2s+1}
(-x)^{s-2a_1} \nonumber \\
&&  \times \sum_{n=-\infty}^{\infty}b^{\nu}_n(-s)
F(-n-\nu-a_1+a_2,n+\nu-a_1+a_2+1;-2a_1+s+1;x) \nonumber \\
\makebox[-3mm]{}&=&\makebox[-3mm]{}
\frac{1}{\tilde A_s} \left(\frac{-x_r}{1-x_r}\right)^{2s} 
\left[\sum_{n=-\infty}^{\infty}a^{\nu}_n(-s)\right]
\left[\sum_{n=-\infty}^{\infty}a^{\nu}_n(s)\right]^{-1}
\left[\frac{r_+ -r'_-}{(r_+ -r_-)(r_- -r'_-)}\right]^{2s} \nonumber \\
&& \times \frac{\Gamma(-2a_1+s+1)}{\Gamma(-2a_1-s+1)}
R^{\nu}_{in;s}\;,
\ena
where we used relations in Eqs.(4.22) and (4.20). 

Thus the T-S identity implies that the following  identity 
among the sums of coefficients should be satisfied, 
\bea
\sum_{n=-\infty}^{\infty}a^{\nu}_n(-s) = |C_s|^2
\left[\frac{a^2}{\alpha(r'_+ -r_+)(r_-  -r'_-)}\right]^{2s}
\left| \frac{\Gamma(\nu+a_1+a_2-s+1)}
{\Gamma(\nu+a_1+a_2+s+1)} \right|^2
\sum_{n=-\infty}^{\infty}a^{\nu}_n(s)
\;.\nonumber\\
\ena

\section{Summary and discussions}

In our previous paper~[8], we showed that both angular and 
radial equations for the Teukolsky equation in Kerr-de Sitter 
(for all massless fields) and Kerr-Newman-de Sitter 
(massless fields except electromagnetic and gravitational 
fields) geometries are transformed to Heun's equation and 
derived  analytic solutions in the form of series of 
hypergeometric functions. The solution of the radial 
equation presented in Ref.8 is convergent for $r<r_+'$ 
with $r_+'$ being the de Sitter horizon. In this paper, 
we constructed the analytic solution of the radial 
equation which is convergent for $r_+<r$ with $r_+$ 
being the outer horizon.  By matching these solutions 
in the region where both solutions converge, {\it i.e.,} 
$r_+<r<r_+'$, we obtained the analytic solution 
valid for the entire region of $r$. 
We showed analytically that our solution satisfies 
the Teukolsky-Starobinsky identities which are quite nontrivial 
differential transformations. 

Although  the use of our solution in this paper is 
less clear than that for the Kerr geometry case, we 
believe that our solution will be 
relevant to the physical world, especially in the early 
stage of the universe where the cosmological constant 
plays important role. By using our solution, 
we can derive the analytic 
expression of the decay rate 
for emission of massless particles from the Kerr-de Sitter 
or the Kerr-Newman-de Sitter black hole~[11], which 
can be used to examine a correspondence between 
quantum gravity on anti-de Sitter space and a conformal field theory 
defined on its boundary~[12,13],

\newpage

\setcounter{section}{0}
\renewcommand{\thesection}{\Alph{section}}
\renewcommand{\theequation}{\thesection .\arabic{equation}}
\newcommand{\apsc}[1]{\stepcounter{section}\noindent
\setcounter{equation}{0}{\Large{\bf{Appendix}}}}

\apsc{Various properties}

\noindent
(a) The proof that $\Delta_r^{-s}R_{-s}^*$ is a solution
 
Form the radial Teukolsky equation for $R_s$, we derive 
the equation for $\Delta_r^s R_s$. We find 
\bea
&&\makebox[-10mm]{}\Bigg\{ \ 
   \Delta_r^s\frac{d}{dr}\Delta_r^{1-s}\frac{d}{dr}
       +\frac{1}{\Delta_r}\left[ 
          (1+\alpha)^2 \left(K-\frac{eQr}{1+\alpha}\right)^2 
  - is(1+\alpha)\left(K-\frac{eQr}{1+\alpha}\right) 
      \frac{d\Delta_r}{dr} \right] \nonumber \\
&& \makebox[-5mm]{}
  +4is(1+\alpha)\omega r -\frac{2\alpha}{a^2}(s-1)(2s-1) r^2
  -s(1-\alpha)-2iseQ -\lambda_s \ \Bigg\}\Delta_r^s R_s(r) = 0.
\ena
By using the relation $\lambda_s=\lambda_{-s}$, we find that 
this is the equation which $(R_{-s})^*$  satisfies. 
By changing $s$ to $-s$, we conclude that $(\Delta_r^{-s}R_{-s})^*$ 
satisfies the same Teukolsky equation as $R_s$ does.   

\vskip 3mm 
\noindent
(b) The proof of $\nu(s)=\nu(-s)$ 
 
With this choice of $A_i$ in Eq.(4.4), coefficients in the 
recurrence relations are expressed by  
\bea
\alpha^{\nu}_n(s) &=& 
-\frac{(n+\nu+a_1-a_2+1)(n+\nu+a_3-a_4+1)
                             |n+\nu+a_1+a_2+s+1|^2}
{2(n+\nu+1)(2n+2\nu+3)}\;,
\nonumber \\
\beta^{\nu}_n(s) &=& \frac{(a_1-a_2)(a_3-a_4)[s^2+(a_1+a_2)(a_3+a_4)]}
{2(n+\nu)(n+\nu+1)}
+\left(\frac{1}{2}-x_r\right)(n+\nu)(n+\nu+1) 
\nonumber\\
&& \makebox[-7mm]{}
+\left[\frac{1}{2}
-\frac{(r_+ +r'_+)^2}{(r_+ -r_-)(r'_+ -r'_-)}\right]s^2
-a_1a_3+a_2a_4+\frac{1}{2}+x_r(a_1^2+a_2^2) 
\nonumber\\
&& \makebox[-7mm]{}  -\frac{2a^4 (1+\alpha)^2}{\alpha^2 {\cal D}}
\frac{(r_- - r'_+)^2 (r_- - r'_-)^2 (r_+ - r'_-) (r'_+ - r'_-)}
{r_+ - r_-}  
\nonumber\\
&& \makebox[-7mm]{}
\times\Bigg\{-\omega^2 r_+^3(r_+ r_- - 2r_- r'_+ + r_+ r'_+)
+2a\omega(a\omega-m)r_+ (r_- r'_+ - r_+^2)  
\nonumber\\
&&  -a^2(a\omega-m)^2(2r_+ - r_- - r'_+) +\frac{eQ}{1+\alpha}
\big[ \ \omega{r_+}^2(r_+ r_- + r_+^2 -3r_- r'_+ + r_+ r'_+)
\nonumber \\
&&  
- a(a\omega-m)(r_+ r_- -3r_+^2 +r_- r'_+ + r_+ r'_+)
\  \big]    
  + \left(\frac{eQ}{1+\alpha}\right)^2 r_+(-r^2_+ +r_- r'_+) \
\Bigg\} 
\nonumber\\
&& \makebox[-5mm]{}
-\left[\frac{2{r'_-}^2}
{(r_+ - r_-)(r'_+ -r'_-)}+x_\infty\right]
-\frac{a^2}{\alpha(r_+ - r_-)(r'_+ - r'_-)} \lambda_s, 
\nonumber\\
\gamma^{\nu}_n(s) &=& -\frac{(n+\nu-a_1+a_2)(n+\nu-a_3+a_4)
                             |n+\nu+a_1+a_2-s|^2}
{2(n+\nu)(2n+2\nu-1)}\;.
\nonumber\\
\ena
And we see that 
\bea
\alpha^{\nu}_{n-1}(s)\gamma^{\nu}_n(s)=
\frac{|n+\nu+a_1-a_2|^2|n+\nu+a_3-a_4|^2|(n+\nu+a_1+a_2)^2-s^2|^2}
{4(n+\nu)^2(2n+2\nu-1)(2n+2\nu+1)},
\ena

Now we observe that the transcendental 
equation which determine $\nu$ given in Eq.(3.6) contains 
coefficients as $\beta_n^\nu(s)$ and a special combination 
$\alpha^{\nu}_{n-1}(s)\gamma^{\nu}_n(s)$. From expressions 
in Eqs.(A.2) and (A.3), we find that both $\beta_n^\nu(s)$ 
and $\alpha^{\nu}_{n-1}(s)\gamma^{\nu}_n(s)$ are 
even functions of $s$. Furthermore, we see that the 
transcendental equation for $\nu$ is an equation of 
real coefficients. Therefore if $\nu(s)$ 
is a solution, then  $\nu(-s)$ is a solution, and also 
$\nu(s)^*$ is a solution. 
By the uniqueness of the solution, we conclude that the solution 
$\nu(s)$ is  an even function and a real function of $s$.


\end{document}